\documentclass[twocolumn,showpacs,preprintnumbers,amsmath,amssymb]{revtex4}

\usepackage{graphicx}% Include figure files
\usepackage{dcolumn}% Align table columns on decimal point
\usepackage{bm}% bold math
\usepackage{color,epsfig,wrapfig,subfigure}
\usepackage{float}
\usepackage{enumerate}

\begin{document}

\title{Total Angular Momentum Conservation During Tunnelling through Semiconductor Barriers}

\author{U. Gennser,$^{(1)}$ M. Scheinert,$^{(2)}$ L. Diehl,$^{(2, 3)}$
S. Tsujino,$^{(2)}$  A. Borak,$^{(2)}$  C. V. Falub,$^{(2)}$ D.
Gr\"{u}tzmacher,$^{(2)}$ A. Weber,$^{(2)}$ D. K. Maude,$^{(4)}$
G. Scalari,$^{(5)}$ Y. Campidelli,$^{(6)}$ O. Kermarrec,$^{(6)}$ and D. Bensahel$^{(6)}$}

\affiliation{$^{(1)}$ CNRS-LPN, Route de Nozay, F-91960 Marcsoussis, France\\
$^{(2)}$ Paul Scherrer Institut, CH-5232 Villigen,
Switzerland\\
$^{(3)}$ Div. of Engineering and
Applied Sciences, Harvard University, Cambridge, MA 02138, USA\\
$^{(4)}$ GHMFL, CNRS, F-38042 Grenoble France\\ 
$^{(5)}$ Universit\'e de Neuch\^atel, CH-2000 Neuch\^atel, Switzerland\\ 
$^{(6)}$ STMicroelectronics, F-38926 Crolles Cedex, France\\ }

\date{\today}

\begin{abstract}

We have investigated the electrical transport through strained
$p-Si/Si_{1-x}Ge_x$ double-barrier resonant tunnelling diodes.
The confinement shift for diodes with different well width,
the shift due to a central potential spike in a well, and magnetotunnelling
spectroscopy
demonstrate that the first two resonances are due to
tunnelling through heavy hole levels, whereas there is no sign of tunnelling through
the first light hole state. This demonstrates for the 
first time the conservation of the total angular momentum in valence band
resonant tunnelling.
It is also shown that conduction through light hole states is possible in many 
structures due to tunnelling of carriers from bulk emitter states.

\end{abstract}

\maketitle

The challenge of introducing spin as an additional degree of
freedom in semiconductor devices has lately attracted great
attention.\cite{Prinz, Wolf} One approach to couple the
spin to the carrier motion is through the spin-orbit
interaction;  one suggestion is to use it in conjunction with resonant
tunnelling devices (RTDs) for injection and detection of spin
currents.\cite{Hall, Glasov}  Whereas the spin-orbit
coupling in the conduction band, mediated by the Dresselhaus
mechanism \cite{Dresselhaus, Malcher} or the Rashba
mechanism,\cite{Bychkov} is generally rather weak, the interaction is strong in the
valence band.  Since this band is made up from p-orbitals, the
interaction term $V_{so} \sim {\bf L}\cdot {\bf S}$ is non-zero,
and there is a strong coupling between the orbital angular
momentum ${\bf L}$ and the spin ${\bf S}$, so that the total
angular momentum $ J = L + S$ is a proper eigenvalue at the band
edge. \cite{luttinger}  In order to examine the feasibility of such devices for
spintronics applications, one may therefore already consider spin
(or $J$) detection in p-RTDs.  It is then rather disconcerting to
find, that in all previous investigations, tunnelling has been observed from heavy
hole states (HH; with $(J, m_{J}) = (3/2, \pm 3/2)$ at $k = 0$)
to light hole states (LH; $(3/2, \pm 1/2)$) or split-off states
(SO, $(1/2, \pm 1/2)$).\cite{Mendez, Liu, Lewis, Hayden, Gennser1}  
It has been proposed
that this non-conservation of the total angular momentum $(J, m_{J})$ in
resonant tunnelling is due to either the band mixing at finite
in-plane momentum $k_{p}$, or because of interface roughness
scattering. However, especially in strained quantum wells, the
non-parabolicity and band mixing for the lowest states is
quite small.  This suggests that scattering plays a large role even
in systems with interfaces known for their good quality.  In our
present study, we show the absence of resonances in the $I-V$
characteristics from heavy holes tunnelling through the first
light-hole state in a double barrier p-type quantum well.  This
demonstrates conclusively that there is ${\bf J}$ conservation during
resonant tunnelling.  Furthermore, by investigating specially
designed RTDs, we are able to show that the emitter structure away
from the barrier interface may explain an apparent mixing of $J$
in the tunnelling process. 

\begin{figure}[h]
    \begin{center}
    \epsfig{file=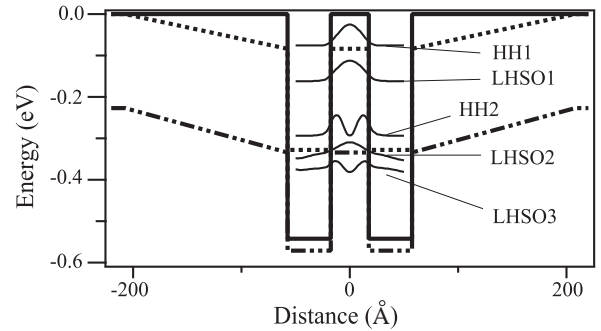, width=78 mm}
    \end{center}
    \caption{\small Schematic Structure of the investigated RTD
    with a $35$\AA $ $ QW indicating the heavy hole valence band edge
    (filled line), light hole valence band edge (dotted line) and split-off band edge
    (dashed-dotted line).}
    \label{struc}
\end{figure}

The samples were grown by molecular beam epitaxy on 
fully relaxed Si$_{0.5}$Ge$_{0.5}$ pseudosubstrates, of which the
top 2 $\mu$m is p-doped, $p = 1\times10^{19}cm^{-3}$. The active part of the initial
structures consist of $40$\AA $ $ barriers surrounding a single Si$_{0.2}$Ge$_{0.8}$ 
quantum well (QW) of width $W$ ($W = $$25$, $35$, or $45$\AA $ $ for three different
samples).  Symmetrically on either side of the active structure are $150$\AA $ $ thick
SiGe emitter layers that are
linearly graded, from $80$\% Ge closest to the barriers to $50$\% Ge away from the barriers.
These emitter layers  consists of an undoped spacer ($100$\AA , closest to the barriers)
and a doped part ($50$\AA, 
$p = 2\times10^{18}cm^{-3}$). 
A $2000$\AA, $p = 2\times10^{18}cm^{-3}$
Si$_{0.5}$Ge$_{0.5}$ top contact layer terminates the structure.
The corresponding structure for the
$35$\AA $ $ QW is schematically shown in Fig. \ref{struc}. For
clarity, the graded emitter region is also shown, and the lowest energy levels
in the quantum well at $k_p = 0$ are indicated.  Due to the strain splitting,
only HH states will be populated in the emitter closest to the barrier.
Since the LH
and SO bands are coupled even for zero in-plane momentum $k_p$, we
have denoted these states as LHSO; however, the LHSO1 level
is in fact predominately LH.

The diodes were processed into mesas with diameters varying
between $10\mu m$ and $300\mu m$. All measurements were performed
at $T \le 4$K, using
separate voltage and current leads connected to both diode
contacts, unless otherwise stated.
However, we found that the resonance voltages changed by less than $10$\% $ $ 
between 4K and 77K.

\begin{figure}[h]
    \begin{center}
  \epsfig{file=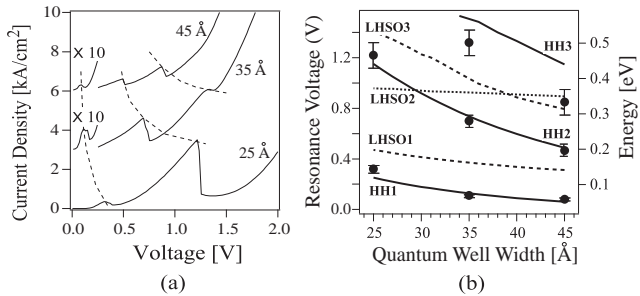,width=85mm}
  \end{center}
  \caption{\small (a) $I-V$ characteristics at $T = $ 77K of the 
     RTDs with a $25$\AA, $35$\AA $ $
     and $45$\AA $ $ quantum well. The curves for the $35$\AA $ $ and $45$\AA $ $ sample
     are shifted
  % $3kA/cm^2$ and $6kA/cm^2$ 
     along the y-axis and magnified
     at the lowest voltages for clarity. The dashed lines are guides to
     the eye to follow the shift of the resonances. (b) Peak voltages vs.
     quantum well thickness (filled dots, left hand scale) and compared with 
     quantum well
     levels calculated for an unbiased QW using a 6-band model 
     (lines - HH states, dotted lines - LHSO states). The error bars indicate the variation 
     between different diodes of the same structure.
     The left and right hand scales relate the energy and voltage to 
     each other through the simple model described in the text.}
  \label{IVpic}
\end{figure}

In Fig. \ref{IVpic}a the 77K current versus voltage characteristics of
the three different RTDs are plotted. They show up to three
resonances, with a maximum peak-to-valley current ratio of $5:1$ at $4$K.
These characteristics are comparable to the best p-type RTDs in
any material system, and indicate a good interface quality
minimizing interface-roughness assisted tunnelling. In
the following we will focus on the two lowest resonances.

The first evidence for assigning the resonances comes from the
confinement shift clearly evident in the $I-V$ characteristics. 
The shift, obtained from measurements of the smallest diodes,
is unaffected by the contact layer resistance from the substrate, as verified by
the dependence of the current on the mesa size.  In
Fig. \ref{IVpic}b the resonance voltages vs. well width are
plotted.  On the right hand scale, these are compared with the
calculated energies. The scales can be directly compared by
assuming a linear voltage drop across the double barriers, the quantum well
and the undoped part of the
structure.
%%INSERT ->
Including the Stark changes the energies much less than the measurement uncertainties.
%%END INSERT  
Because of the graded nature of the emitter, the zero bias emitter states
lie $\approx$30 meV higher then than the quantum well edge.  This is 
included as an energy offset between the two scales.
The so-called 'lever arm' - i.e. the ratio between the energy drop between the emitter 
and the centre of the
quantum well and the applied voltage - is in good agreement with what can be expected from
geometrical considerations, and the energies are consistent with those obtained from
intersubband absorption measurements.\cite{Diehl}
A  good agreement between theory and experiment is found if
the first two resonances correspond to tunnelling through the HH1
and HH2 states, respectively. Moreover, the
difference between the first two resonances increases with decreasing
well width. This effect is only obtained for 
states with different index, such as HH1 and HH2. 
% However, in view of the simplicity of
%the model - e.g., neither depletion width nor carrier
%accumulation in the structure is taken into account - the
%confinement shift can only be taken as an indication of the nature
%of the resonances.  In the following, we describe a comparison
%with wells having a potential spike in the middle, as well as
%magnetotunnelling experiments, which give conclusive evidence that
%the above assignment is correct.
%% INSERT ->

In view of the simplicity of the model - e.g. neither depletion width nor carrier 
accumulation in the structure is taken into account - this result alone can only be
taken as an indication of the nature of the resonances.  
However, support for the model is found through magnetocurrent oscillations.
For low, fixed $V$
and with a magnetic
field $B$ applied parallel to the current, it is possible to observe weak
oscillations periodic in $1/B$ (period $B_{f}$).  
They are due to Landau levels passing
through the quasi-Fermi energy in the emitter accumulation layer, the two-dimensional
charge density of which is $p_{e} = 2eB_{f}/h$.  Unlike similar
oscillations in GaAs/AlAs p-type RTDs \cite{Hayden2}, 
no decrease in $B_{f}$ is found as $V$ passes
through the resonances, from which we conclude that the charge density in the quantum 
wells is negligible.  Furthermore, the
electric field $F = ep_{e}/\epsilon \epsilon_{0}$ over the QW structure is in 
reasonable agreement with the simple lever arm model.

\begin{figure}[h]
    \begin{center}
  \epsfig{file=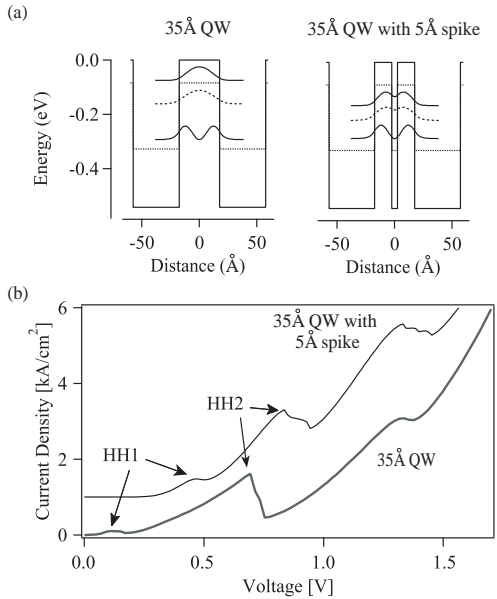,width=65mm}
  \end{center}
    \caption{\small (a) Schematic band diagrams of the $35$\AA $ $ Si$_{0.2}$Ge$_{0.8}$
    quantum well
    without and with a central $5$\AA $ $ Si barrier, showing the HH and LH potential,
    and  HH1,
    LHSO1 and HH2 wave functions. Only the HH2 state remains almost unaffected by 
    the potential spike.
    (b) 77K $I-V$ characteristics for samples with (shifted up for clarity) 
    and without a central
    $35$\AA $ $ barrier. The resonances corresponding to tunnelling through 
    the HH1 and HH2 states are indicated with arrows.}
    \label{spike}
\end{figure}

Further, conclusive evidence that the above assignment of the resonances is correct
can be found in experiments where the resonances are shifted by a central
potential spike.
%% <- END INSERT 
Even symmetry states (HH1, LHSO1), with a wave function maximum in
the middle of the quantum well, are much more affected by a
central, repulsive potential spike than odd symmetry states (HH2,
LHSO2).\cite{JYM}   In our samples, the spike has been approximated by a thin
Si layer; a $35$\AA $ $ QW with a $5$\AA $ $ spike in the middle
was investigated and compared to the initial $35$\AA $ $ structure
(Fig. \ref{spike}a). The plotted wavefunctions 
in the figure give a clear picture of the described effect.
An example of the $I-V$ characteristics measured at 77K is displayed in Fig.
\ref{spike}b) and clearly demonstrate the predicted behaviour for
the HH1 and HH2 states. 
We find shifts for the first and second resonances equal to $(0.31 \pm 0.03) V$ and
$(0.06 \pm 0.08) V$, respectively, where the uncertainty is due to the 
natural scatter of the measured resonances
for different diodes of the same structure.  The values compares well with the 
calculated values (using the model described above, including Stark shifts) of $0.21 V$ 
and $0.06 V$ for the HH1 and HH2 resonance respectively.\cite{HH1}  
In contrast, assuming 
a lever arm compatible with the second resonance due to LHSO1 tunnelling, 
the expected shift 
would be $\ge 0.2 V$.

\begin{figure}[h]
    \begin{center}
  \epsfig{file=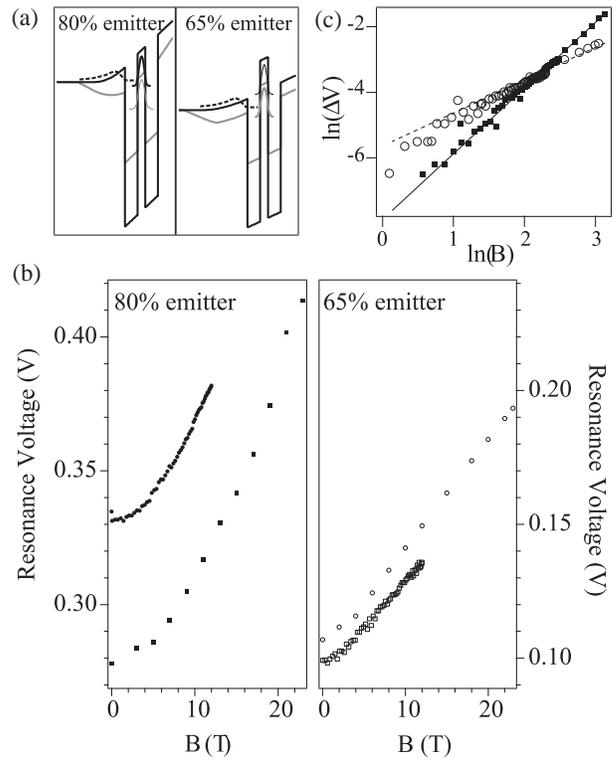,width=80 mm}
  \end{center}
  \caption{\small (a) The schematic band diagram of the $25$\AA $ $ quantum well
  sample with an $80\%$ emitter and a $65\%$ emitter at 
  biases close to their respective resonances.
  %a bias of 255 mV and 145 mV, 
  %respectively.  
  The HH potential (thick black line), the LH potential (thick grey), 
  the quantum well HH1 (bold) and LHSO1 (grey) wavefunctions as well 
  as the confined emitter state wave function (dashed) are shown.
  (b) The effect of a magnetic field parallel to the
  interface on the resonance voltages for the sample with a $80\%$
  and one with a $65\%$ emitter. Two different diodes for each type of sample were measured
  up to $12$T and $23$T respectively, each with slightly different resonance voltage but 
  with identical magnetic field behaviour.
  (c) Log of the resonance voltage difference $\Delta V = V(B)-V(B=0)$ vs. log of the
  magnetic field parallel to the interface, for the $80\%$ (filled dots) and $65\%$ 
  emitter sample (open dots).  Lines $d(ln(\Delta V))/d(lnB) = 1$ and 
   $d(ln(\Delta V))/d(lnB) = 2$ are shown as guides for the eye.}
  \label{LHpic}
  \end{figure}

Having shown that it is possible to observe $J$-conservation in
these tunnelling experiments, we now try to understand the
difference between the present samples and those of previous
studies, where tunnelling through LHSO states was observed.  One
important contrast is the higher strain used in the
present study.  For example, in previous studies of Si/SiGe RTDs
on Si substrates, the Ge content was around $20 - 25 \%$. \cite{Liu, Gennser1}¥
One consequence is that the HH and LHSO states in the emitter were
less decoupled in these samples, with a separation between the HH and LH potentials
$\le 45$ meV, whereas for the present samples it is $\approx 85$ meV.
To study the role of the emitter, a
structure with a $25$\AA $ $ QW and an emitter region with a grading
from $50$\% $ $ to $65$\% $ $  was investigated (See Fig. \ref{LHpic}(a)). 
Two resonances, at $\approx$100 mV and $\approx$470 mV, are observed in
this 'emitter ramp' sample. 
% It is designed so that the HH1 should lie below the 
%band edge of the emitter (the 'ramp') at
%zero bias, but because of the uncertainty in the band offsets, it is not possible to
%exclude directly that the first resonance is due to tunnelling through HH1.
%However, for reasonable values of the lever arm,
%the small voltage difference between the two resonances (370 mV) compared
%to the calculated energy difference between the HH1 and HH2 states (320 meV, see Fig. 2)
%rules out the possibility that we observe the usual HH1 and HH2 resonances in this sample.
The second resonance voltage is compatible with the estimated resonance 
voltage of the tunneling through HH2 but the first resonance is likely due 
to the tunneling through LHSO1: the tunneling through HH1 is prohibited by design. 
Also the 370 mV separation between the two resonances is more than a factor of 2 
smaller than the separation between the HH 1 and the HH2 resonances of  
$80$\% $ $ emitter sample (Fig.2).

%% INSERT ->
To further compare these resonances, we use magnetotunnelling spectroscopy
with $B$ up to 23 T.
A magnetic field $B_{\perp}$
applied perpendicular to the current $I$ accelerates the carriers in the
direction perpendicular to both $B_{\perp}$ and 
$I$, so that they tunnel through the quantum well levels
at a non-zero in-plane momentum, centered around $\Delta k_{p} =
q \Delta s B_{\perp}/\hbar$ where $\Delta s$ is the tunnelling
distance.\cite{Hayden} The
in-plane dispersion relations can then be mapped out and compared with the
calculated dispersion $E(k_{p})$.\cite{Hayden, Hayden2, Gennser1, Gassot} 
All the HH1 resonances of the three regular structures show a parabolic
behaviour. The corresponding effective masses
(0.04 $m_0$, 0.15 $m_0$, and 0.13 $m_0$ for
the $25$\AA, $35$\AA $ $ and $45$\AA $ $ wells, respectively)
are in reasonable agreement with the calculated dispersions 
(0.17 $m_0$, 0.155 $m_0$, and 0.144 $m_0$) though
quantitative comparisons are difficult to make. \cite{Hayden2}
%% <- END INSERT 

In Fig. \ref{LHpic}(b) we compare the  $B_{\perp}$
shift for the first resonance of the sample with a $25$\AA $ $ QW and an
80$\%$ emitter and of the emitter ramp sample. 
%The 80$\%$ emitter resonance shows a parabolic 
%behaviour, as is the case with all other resonances studied in this work, 
%the second resonance
%of the emitter ramp sample included.  The exception is the first resonance
%of this sample, which shows a very distinct linear behaviour.  
%% INSERT ->
In contrast to the 80$\%$ emitter resonance, the first resonance of the 
emitter ramp sample shows a very distinct linear behaviour.
A log plot clearly demonstrates these dependences (Fig. \ref{LHpic}c). This 
indicates that it is the first resonance rather than the second that is not due to
tunnelling through one of the HH states.  Furthermore, a magnetic field  $B_{\perp}$
cannot lead to a linear energy shift of the valence band QW states or the emitter 
states next to the barrier.
The
Zeeman effect is given by $E_{Z} = \kappa \mu _{B}¥{\bf J} \cdot {\bf B}$
(plus a small term proportional to $B^{2}$) \cite{Luttinger}, which
is only a small
perturbation since the direction of $J$ is frozen in the direction of the
confinement and the strain.  Since the well thickness is much smaller than the cyclotron
orbit even for the highest fields, Landau level formation can also
be excluded.  Neither can the linear shift in  Fig. \ref{LHpic} be explained
by the acceleration in k-space, since the levels are quite parabolic, and
never linear in $k_p$. In fact, we find that only an unstrained 
valence band bulk state can give rise to the observed
linear shift.
%In order to further elucidate the nature of this LHSO1
%resonance the resonance voltages were studied as a function of
%$B_{\perp}$ (Fig. \ref{LHpic}). A very different
%behaviour for the lowest resonance of the two samples is observed.
%While it has a quadratic $B_{\perp}$-dependence in the case of the
%$80\%$ emitter sample, it is linear for the $65\%$ emitter sample.
%A log plot clearly demonstrates these dependences (Fig. \ref{LHpic}c). 
%The significance of such
%a linear dependence becomes clear when considering the different
%magnetic field contributions to a p-type quantum well level, none
%of which is linear in $B$.  The acceleration in k-space has
%already been discussed; since the levels are quite parabolic, and
%never linear in $k_p$, this cannot result in the observed
%behaviour.  Since the well thickness is much smaller than the cyclotron
%orbit even for the highest fields, Landau level formation can also
%be excluded.  Finally, we note that in the valence band the
%Zeeman effect is given by $E_{Z} = \kappa {\bf J} \cdot {\bf B}$
%(plus a small term proportional to $B^{2}$).\cite{Luttinger}
%However, the direction of $J$ is frozen in the direction of the
%confinement and the strain; the Zeeman effect is only a small
%perturbation as long as the magnetic energy is smaller than the
%energy level separation in the wells.  This leads to the
%conclusion that only a valence band bulk state can result in a
%linear shift in energy. 
We propose that there are two reservoires of holes in
the emitter: states confined close to the Si barrier and states in the unstrained 'bulk'
part of the emitter.  The latter, tunnelling through the LHSO1 state,
are responsible for the first resonance of the ramp
emitter sample.  In the bulk the $\bf{J}$ vector is free to turn along
the B-field axis, and with $\bf{J}$ perpendicular to the growth
axis, the quantum well state will 'see' a mixed HH-LHSO state
coming from the emitter. Because of the lower Ge content in this
emitter, the barrier for the holes from the bulk is smaller,
making it possible for them either to tunnel directly into the
quantum well states, or to form hybrid states with the emitter
states in the HH emitter well. The Landau level separation in the
Si$_{0.5}$Ge$_{0.5}$ bulk is $\approx$ 0.6 meV/T, and the Zeeman
energy a factor of 2-10 smaller.\cite{Winkler} 
This compares reasonably well with the   
measured slope of 4.1 mV/T
$\approx$ 1.4 meV/T.  It seems plausible that the apparent
tunnelling from HH to LHSO states in other p-type RTDs may be due
to the inevitable bulk part of the emitter, as well as band mixing
in the well states.  A similar linear behaviour has indeed been
observed in a Si/Si$_{0.75}$Ge$_{0.25}$ RTD with the strain
fully in the SiGe layer. \cite{Gassot}

Concerning the third resonance of the $35$\AA $ $ and $45$\AA $ $
sample, the fit with the
energy levels in Fig. \ref{IVpic} indicates that it
corresponds to tunnelling through the second LH-like state 
(LHSO3 in the figure), and this also 
agrees with the observed shift in the sample with a central Si spike. This state is
much less parabolic than the three lower states, and one would
therefore expect a larger amount of band mixing.  However, 
further experiments are necessary to confirm this.

We have demonstrated that the total angular momentum is conserved
during resonant tunnelling in a system with strong spin-orbit
coupling.  This does not necessarily imply that the same holds
true for the case of weakly coupled spin, but is certainly an
encouraging sign.  However, it may also have direct implications
for the field of spintronics, since in order to inject spin in a
semiconductor, a possible path is through the growth of magnetic
semiconductors as electrodes. Much of the work has been focused on
GaMnAs alloys, where the Mn not only provides the ferromagnetic
properties, but also is a p-dopant.\cite{Dietl, Mattana}
Finally, it should also be noted that these results may have an
additional relevance for the development of a Si/SiGe based
quantum cascade laser, in they exclude one of the possible
non-radiative conduction paths
for the HH carriers in these structures.\cite{Gennser2}

\begin{acknowledgments}
We would like to thank J. Faist for help with this work. It
has been partly financially supported by the Swiss National Science
Foundation, the EC Contract Si-GeNET, "R\'egion Ile de
France", "Conseil G\'en\'eral de l'Essonne" and the program 
Nano2008.

\end{acknowledgments}


\begin{thebibliography}{}
\bibitem{HH1} The larger discrepancy for the HH1 is expected since it is more
sensitive to the exact thickness of the Si spike.
\end{thebibliography}
\end{document}